\documentstyle[twoside,fleqn,espcrc2]{article}

\newcommand{\AmS}{{\protect\the\textfont2
  A\kern-.1667em\lower.5ex\hbox{M}\kern-.125emS}}
\newfam\scrfam
\batchmode\font\tenscr=rsfs10 \errorstopmode
\ifx\tenscr\nullfont
\else	\font\sevenscr=rsfs7 
	\font\fivescr=rsfs5 
	\skewchar\tenscr='177 \skewchar\sevenscr='177 \skewchar\fivescr='177
	\textfont\scrfam=\tenscr \scriptfont\scrfam=\sevenscr
	\scriptscriptfont\scrfam=\fivescr
	\def\scr{\fam\scrfam}
	\def\cal{\scr}
\fi
\newfam\msbfam
\batchmode\font\twelvemsb=msbm10 scaled\magstep1 \errorstopmode
\ifx\twelvemsb\nullfont\def\Bbb{\bf}
\else	\catcode`\@=11
	\font\tenmsb=msbm10 \font\sevenmsb=msbm7 \font\fivemsb=msbm5
	\textfont\msbfam=\tenmsb
	\scriptfont\msbfam=\sevenmsb \scriptscriptfont\msbfam=\fivemsb
	\def\Bbb{\relax\expandafter\Bbb@}
	\def\Bbb@#1{{\Bbb@@{#1}}}
	\def\Bbb@@#1{\fam\msbfam\relax#1}
	\catcode`\@=\active
\fi
\font\eightrm=cmr8		\def\xrm{\eightrm}
		
\font\eightit=cmti8		\def\xit{\eightit}

\font\tentt=cmtt10
\font\tenrm=cmr10

\font\fiverm=cmr5		\def\trm{\fiverm}

\def\/{\over}
\def\*{\partial}
\def\a{\alpha}
\def\b{\beta}
\def\c{\gamma}
\def\d{\delta}
\def\e{\varepsilon}
\def\fun{\varphi}

\def\g{\gamma}
\def\k{\kappa}
\def\l{\lambda}
\def\p{\phi}

\def\t{\theta}

\def\x{\xi}

\def\D{\Delta}
\def\F{{\cal F}}
\def\G{\Gamma}
\def\L{{\cal L}}
\def\LL{\Lambda}

\def\punkt{\,\,.}
\def\komma{\,\,,}
\def\.{.\hskip-1pt }
\def\is{\!=\!}
\def\-{\!-\!}
\def\+{\!+\!}
\def\={\!=\!}
\def\>{\!>\!}

\def\half{{\lower1.5pt\hbox{\eightrm 1}\/\raise1.5pt\hbox{\eightrm 2}}}
\def\frac#1{{\lower1.5pt\hbox{\eightrm1}\/\raise1.5pt\hbox{\eightrm#1}}}
\def\tfrac#1{{\lower1.5pt\hbox{\trm1}\/\raise1.5pt\hbox{\trm#1}}}
\def\genfrac#1#2{{\lower1.5pt\hbox{\eightrm#1}\/\raise1.5pt\hbox{\eightrm#2}}}
\def\quarter{\frac{4}}

\def\tr{\hbox{\rm tr}}
\def\ttr{\hbox{\xrm tr}}
\def\ie{{i.e.,}}

\def\DBI{\hbox{Dirac--Born--Infeld}}
\def\WZ{\hbox{Wess--Zumino}}
\def\dbi{{\hbox{\fiverm DBI}}}
\def\wz{{\hbox{\fiverm WZ}}}
\def\det{\hbox{\rm det}}
\def\II{I\hskip-0.6pt I}
\def\one{1\hskip-3.85pt1}
\def\id{\one}
\def\gel{\g_{1\hskip-.5pt 1}}

\def\be{\begin{equation}}
\def\ee{\end{equation}}
\def\beq{\begin{eqnarray}}
\def\eeq{\end{eqnarray}}

\title{\vbox{\vskip-24pt\hbox to\hsize{\tenrm\hfill G\"oteborg-ITP-96-20}
		\vskip-8pt
		\hbox to\hsize{\hfill \tentt hep-th/9612153}\vskip12pt}
Aspects of D-brane actions}

\author{Martin Cederwall
	\address{Institute for Theoretical Physics\\
	G\"oteborg University and Chalmers University of Technology\\ 
        S-412 96, G\"oteborg, Sweden}%
}

\begin{document}

\begin{abstract}
A couple of issues concerning the effective dynamics of D-branes in string
theory are discussed. Primarily, I am concerned with linearization of the
actions by introduction of non-propagating fields and a full super- 
and $\k$-symmetric description of D-branes. This contribution summarizes
the work of refs.~\cite{CvGMNW,CvGNW,CvGNSW}.
\end{abstract}

\maketitle

\section{BACKGROUND}
D-branes, or Dirichlet branes, play a central r\^ole in non-perturbative
superstring theory \cite{Polchinski,PolchinskiWitten,WittenIV}. 
They occur as soliton-like solutions of the 
low-energy effective actions for the massless sector of type \II\ string
theories, and provide hypersurfaces of Dirichlet boundary conditions
for open strings. D-branes are the objects in string theory carrying
Ramond--Ramond (RR) charges, \ie\ they couple to the antisymmetric
tensor fields in the massless RR sector of type \II\ string theory. 
We call the D-brane configurations ``soliton-like'' because their
r\^ole seems to be intermediate between fundamental and truly solitonic
excitations in that they do not act as fundamental excitation in any
dual picture of the theory.

One part of the effective actions for D-branes, namely the ``kinetic term'',
containing the couplings to the massless fields in the NS--NS sector,
has been known for quite some time, and was obtained by demanding that
the $\b$-functions corresponding to these fields for an open string
ending on a D-brane vanishes \cite{Leigh}. Boundary contributions are cancelled
by the classical variation of the effective D-brane action, which
turns out to be of \DBI\ (DBI) form:
\begin{equation}
I_{\dbi}=-\int d^{p+1}\x e^{-\phi}\sqrt{-\det(g_{ij}+\F_{ij})}\punkt
\end{equation}
This action contains the dilaton $\phi$, the target space metric $g_{mn}$
via its pullback to the world-volume $g_{ij}$ and the antisymmetric
tensor field $B_{mn}$ via $\F_{ij}={\a'\/2\pi}F_{ij}-B_{ij}$. 
The bosonic world-volume
degrees of freedom are thus the embedding together with a world-volume
U(1) potential $A$, $F=dA$. At this point, D-branes differ from ordinary
$p$-branes in that there are bosonic world-volume fields in non-scalar
representations, a fact that becomes important when one wants to
write down supersymmetric actions.
The DBI action thus contains the kinetic part of the action for the
bosonic degrees of freedom, as well as the coupling to the massless
fields in the NS--NS sector.

Another piece of information concerning the actions is the form of
the coupling to the RR antisymmetric tensor potentials. To its full extent,
this was first formulated in refs.~\cite{Douglas,GHT}. To the DBI action should
be added a Wess-Zumino (WZ) term of the form
\begin{equation}
I_\wz=\int e^\F C\komma
\end{equation}
where C is the cochain of RR potentials together with the potentials 
for the dual field-strengths. The products (including the exponential)
are understood to be wedge products of forms. Since $\F$ is a two-form,
we see that D$p$-branes of odd $p$ couple to even potentials, \ie\ those
of type \II B superstring theory, while those of even $p$ couple to
the odd potentials of type \II A. This of course agrees with the
supergravities in which these branes are found as solutions.

Both the formulation of a supersymmetric action and considerations concerning
duality properties \cite{Tseytlin} demand a delicate interplay between the DBI
and WZ parts of the actions.

So, while there has been quite some information about the effective 
dynamics of D-branes, and about their r\^ole in non-perturbative string
theory, important pieces of information have been missing. 
In this contribution, I would like to address two issues.
The first of these is the question of whether it is possible to rewrite
the DBI action in a more tractable form, using auxiliary non-propagating
fields on the world-volume \cite{CvGMNW}. In string theory, the corresponding
procedure takes us from the Nambu--Goto action to the Brink--DiVecchia--Howe
action \cite{BDH}, which
is much more useful in that the coordinates simply become free scalars.
We do not expect the same dramatic simplification as in string theory,
of course, since an auxiliary world-volume metric cannot be gauge-fixed
to the same extent as in the case $p=1$, but such a formulation might
still be illuminating, especially with respect to symmetries.
The second issue is the inclusion of fermionic degrees of freedom
on the world-volume \cite{CvGNW,CvGNSW}. 
We know that D-brane configurations are BPS states, 
conserving half of the supersymmetry. In order to address this question,
we need to understand the mechanism of $\k$-symmetry for D-branes.
This fermionic gauge symmetry removes half of the fermionic variables
on the world-volume, so that the physical ones may generate the appropriate
number of states for a BPS-saturated multiplet. 

At the time of the conference, there were only some partial results
on supersymmetric D-branes. Some of the key ideas used in 
refs.~\cite{CvGNW,CvGNSW}
were developed during this meeting, and since one of the main points
of the talk was supersymmetrization, I found it appropriate rather to
include results obtained after the date of the talk than to pretend
that I still do not know them. Three other papers 
\cite{Aganagic,BergshoeffTownsend,AganagicII} 
by other authors have also appeared since, 
that derive parts of the results of refs.~\cite{CvGNW,CvGNSW}.

I want to take the opportunity to thank the organizers of the 30th
Ahrenshoop symposium for an excellent conference, and my collaborators
Alexander von Gussich, Aleksandar Mikovi\'c, Bengt E.~W.~Nilsson, Per Sundell
and Anders Westerberg, whose joint work I report on here.

\section{``LINEARIZED'' ACTIONS}

In string theory, it has been extremely useful to work with a formulation
where the area action given in terms of the pullback of a background
metric, \ie\ the Nambu--Goto action
\begin{equation}
I_{\hbox{\fiverm NG}}=-\int d^2\xi\sqrt{-\det g_{ij}}\komma
\end{equation}
is replaced by an equivalent action containing an auxiliary 
non-propagating world-sheet metric $\g$,
\begin{equation}
I=-\half\int d^2\xi\sqrt{-\g}\g^{ij}\*_iX^m\*_jX^ng_{mn}\punkt
\end{equation}
The embedding coordinates become ordinary free scalars on world-sheet,
and this allows for the extremely powerful machinery of conformal field
theory. Although the possible gauge choices for higher-dimensional
objects are not expected to lead to as dramatic simplifications as for
strings, it might still be relevant to ask whether a similar reformulation
is possible. A formulation where the dynamical degrees of freedom behave
as free fields would in many respects, such as supersymmetrization,
be preferable.
We refer to action containing the auxiliary metric as
Brink--DiVecchia--Howe--Tucker (BDHT) type actions \cite{BDH,Tucker}.

It is well known that any $p$-brane involving no other bosonic
world-volume degrees of freedom than scalars allows for such a formulation.
It is a straightforward generalization of the string action, namely
\be
\vbox{
\hbox{$I=-\half\int d^{p+1}\xi\sqrt{-\g}\bigl\{\g^{ij}\*_iX^m\*_jX^ng_{mn}
     \bigr.$\hfill}
\hbox{$\phantom{I=-\half\int d^{p+1}\xi\bigl\{}\bigl.
	-(p-1)\bigr\}\komma$\hfill}
}
\ee
Elimination of $\g$ via its (algebraic) equations
of motion yields the equivalent action
\be
I=-\int d^{p+1}\xi\sqrt{-\det g_{ij}}\punkt
\ee
where $g$ is the pullback of the background metric, 
$g_{ij}=\*_iX^m\*_jX^ng_{mn}$. The solution of the equations of motion 
for $\g$ is simply $\g=g$.

In ref.~\cite{CvGMNW}, we investigate the generalization of this procedure
to D-branes, where as mentioned one has a world-volume vector potential
in addition to the embedding coordinates. The hope was that, by the
introduction of an auxiliary world-volume metric, the \DBI\ action 
would become more ``linear'', and that this would be the appropriate
setting for addressing issues like supersymmetrization.
The basic assumption that was made was that the embedding coordinates
should continue to behave as free fields on the world-volume.
We were encouraged by the already known fact \cite{Townsend} that for $p=2$
the equivalent BDHT-type action just includes a quadratic term in
the world-volume field strength. The manipulations in this section 
will be restricted to the \DBI\ part of
the D-brane actions only. Also, since the background fields do not take
part in the procedure, they can be considered just as constants during
this procedure, and nothing is lost by restriction to the flat case.

To this end, one can make a completely general Ansatz for the BDHT lagrangian:
\be
\vbox{
\hbox{$\L=-\half\sqrt{-\g}\,\bigl\{\tr(\g^{-1} g)+\fun(\g^{-1}F)\bigr.$\hfill}
\hbox{$\phantom{\L=-\half\sqrt{-\g}\,\bigl\{\bigr.}\bigl.
	-(p-1)\bigr\}\punkt$\hfill}
}
\ee
Notice that the metric $\g$ only occurs algebraically, so that one in
principle may solve its equations of motion and plug the solution back
into the action. The criterion is then that this procedure should
yield the \DBI\ action, and equivalence is proven.

In order to deal with the algebra associated with solving these equations,
it is convenient to define a set of matrices
\be
{u^i}_j=\g^{ik}g_{kj}\komma\quad
{X^i}_j=g^{ik}F_{kj}\komma
\ee
or, in shorthand,
\be
u=\g^{-1}g\komma\quad
X=g^{-1}F\punkt
\ee
When all relevant entities are expressed in terms of matrices with one
upper and one lower index, all algebra reduces to matrix algebra.
Solving for $\g$ in terms of $g$ and $F$ is equivalent to solving for
$u$ in terms of $X$.


As a warm-up exercise, one may treat the case $F=0$. Then the equations
of motion for $\g^{-1}$ read
\be
0=-\half\tr u+u+{p-1\/2}
\ee
and
the solution is obviously $u=\one$. Plugged back into the action, it
gives the Nambu--Goto action (3), as promised.

In order to incorporate the tensor field, and to treat all values of $p$,
we make the following observation. Due to the antisymmetry of $F$, the
number of independent scalar invariants equals the rank of the world-volume
Lorentz group, \ie\ $n$ for $p=2n-1$ and $p=2n$. One such set of invariants
consists of $\tr X^{2k}$, $k=1\ldots n$, and, for odd $p$, $\det X$.
There is also a matrix identity, expressing $X^{p+1}$ in terms of 
lower powers of $X$ and these invariants. Due to symmetry properties,
the solution of $u$ in terms of $X$ only contains even powers, and
a general Ansatz may be written as $u(X)=\sum_{k=0}^n u_kX^{2k}$, where
the $u_k$'s are scalar functions of the invariants.

An even more useful basis is one where the matrix $X^2$ is diagonalized.
The general structure is that for $p=2n$ there will be $n$ non-zero
eigenvalues with multiplicity two, and a non-degenerate
zero eigenvalue, while for
$p=2n-1$ there will be $n$ non-zero
eigenvalues with multiplicity two. When one restricts
to the subspace spanned by the even powers of $X$, the degeneracy
disappears. The ``eigenvectors'' 
$v_0$, $\{v_i\}_{i=1}^n$ are conveniently
normalized so that they become projection operators on the appropriate
linear subspaces. The expansion we will use is
\be
u(X)=u_0v_0+\sum_{i=1}^nu_iv_i\komma
\ee
where the $u_i$'s are scalar functions of the invariant eigenvalues
$\{\l_i\}$ and the first term is present only for even values of $p$.

The observation that simplifies the calculations in this basis is that
the unknown function $\fun$ has the argument $uX$, and this only enters
with even powers. Any solution $u(X)$ commutes with $X$, so $\fun$
only depends on the combination $X^2u^2=\sum_{i=1}^n\l_i{u_i}^2v_i$,
so that it is a function of the scalars $\{\l_i{u_i}^2\}_{i=1}^n$.
It is then straightforward work to write down the equations of motion
for the auxiliary metric, and the condition that resubstituting its
solution yield the \DBI\ action. These conditions together become:
\be
\vbox to160pt{
\hbox{\underbar{$p$ odd:}\hfill}
\vfill
\hbox{$0=n\-1-\sum\limits_{j\neq i}u_j-\half\fun
	+\half u_i{\*\fun\/\*u_i}\komma$\hfill}
\vfill
\hbox{$\qquad\qquad\qquad\qquad i\=1,\ldots,n\komma$\hfill}
\vfill
\hbox{$\sum\limits_iu_i+\half\fun-(n\-1)
	=\prod\limits_i(1\-\l_i)^{1/2}u_i\punkt$\hfill}
\hbox{\hfill}
\hbox{\underbar{$p$ even:}\hfill}
\vfill
\hbox{$u_0=-(2n\-1)+2\sum\limits_iu_i+\fun\komma$\hfill}
\vfill
\hbox{$0=-u_0+u_i+\half u_i{\*\fun\/\*u_i}\komma$\hfill}
\vfill
\hbox{$\qquad\qquad\qquad\qquad i\=1,\ldots,n\komma$\hfill}
\vfill
\hbox{$u_0=\prod\limits_i(1\-\l_i)u_i^2\punkt$\hfill}
}
\ee
To arrive at these equations, we have used the properties of the 
eigenvectors/projection operators together with the determinant
relations $\det u=u_0\prod_{i=1}^n{u_i}^2$ (first factor only present
for even $p$), 
$\det(1+X)=\prod_{i=1}^n(1-\l_i)$.

At this point, one would like to look for solutions to these equations.
Such solutions provide the BDHT actions, as well as explicit relations
between the auxiliary and induces metrics. The solution becomes
increasingly difficult with increasing world-volume dimensionality.
We will illustrate by starting with the membrane and moving up in $p$.

The first issue is the (already well known \cite{Townsend}) D2-brane. 
$X^2$ is readily diagonalized with
\be
\vbox to27pt{
\hbox{$\l_0=0\komma\quad v_0=\one-{2\/\ttr X^2}X^2\komma$\hfill}
\vfill
\hbox{$\l_1=\half\tr X^2\komma\quad v_1={2\/\ttr X^2}X^2\punkt$\hfill}
}
\ee
(there is no reason to worry about objects that are singular as 
$F\rightarrow0$, they will not occur in the final expressions).
When this is inserted in eq.~(12), the solution is easily found to be
\be
u_0=1-\l_1\komma\quad u_1=1\komma\quad\fun=-\l_1{u_1}^2\komma
\ee
or in terms of the variables of the original actions,
\be
\vbox to 27pt{
\hbox{$\g^{-1}g=\one-\half\tr(g^{-1}F)+(g^{-1}F)^2\komma$\hfill}
\vfill
\hbox{$\fun=-\half\tr(\g^{-1}F)^2\punkt$\hfill}
}
\ee
Thus, the very non-linear \DBI\ action becomes the ordinary action for
a free Max- well field when the auxiliary metric is introduced. 

For the three-brane, the diagonalization yields
\be
\vbox to32pt{
\hbox{$\l_\pm={1\/4}\tr X^2\pm\sqrt{-\D}\komma$\hfill}
\vfill
\hbox{$v_\pm=\half\left[\,1\pm{1\/\sqrt{-\D}}\Bigl(X^2-{1\/4}\tr X^2\Bigr)
	\right]\komma$\hfill}
}
\ee
where $\D(X)=\det X-\genfrac{1}{16}(\tr X^2)^2$.
Here, it takes some guesswork to find the solution, and we refer to
ref.~\cite{CvGMNW} for details. It is
\be
\vbox to 75pt{
\hbox{$u_\pm=\sqrt{1-\l_\mp\/1-\l_\pm}\komma$\hfill}
\vfill
\hbox{$\fun=2\biggl\{1-\sqrt{\bigl(1+\l_+^{\phantom{2}}u_+^2\bigr)
		\bigl(1+\l_-^{\phantom{2}}u_-^2\bigr)}\biggr\}$\hfill}
\vfill
\hbox{$\phantom{\fun}=2\biggl\{1-\sqrt{1+\half\tr(uX)^2
	+\det(uX)}\biggr\}\punkt$\hfill}
}
\ee
The situation is clearly more non-linear than for the membrane, and 
the hope that the presence of an auxiliary metric would simplify the
action does not seem to be fulfilled. One may make two curious remarks
here. The first one is that although the induced and auxiliary metrics
are related via an equation that is non-polynomial in the field strength,
their determinants coincide (this follows from $u_+u_-=1$). We have
not really understood why this should happen, and it may be a clue to
some interesting structure. The second one is that if $\Delta=0$, the entire
non-linearity dissolves, and one again gets the free Maxwell action.
This observation might be interesting for branes with signature (2,2),
where such a constraint is equivalent to selfduality or anti-selfduality
(in signature (1,3) it is too strong, and makes all the $F$ contributions
to the action vanish).
It should also be pointed out that the occurrence of a square root is not
directly related to the square root in the \DBI\ action, but rather 
peculiar to $p=3$.

For the case $p=4$, we are in the peculiar situation that we have been
able to derive the solution of the field equations for $\g$, but not
a closed form for the action they descend from. The non-zero eigenvalues
are $\l_\pm=\quarter\tr X^2\pm[\,\quarter\tr X^4\-{1\/16}(\tr X^2)^2\,]^{1/2}$,
and the relation between the auxiliary and the induced metric is given
by $u_\pm=(1\-\l_\pm)^{-2/3}\,(1\-\l_\mp)^{1/3}$. The function $\fun$ has
as its arguments $t_\pm\is\l_\pm u_\pm^2$, and it is given implicitly by
\be
\vbox to27pt{
\hbox{$t_\pm=u_\pm^2-{1\/u_\mp}\komma$\hfill}
\vfill 
\hbox{$\fun=3+{1\/u_+u_-}-2\,(u_+\+u_-)\punkt$\hfill}
}
\ee
Ironically enough, elimination of $u_\pm$ from these equations amounts
to solving a fifth order equation.

To conclude this section, we have devised a systematic procedure for
introducing the auxiliary metric and writing down the BDHT action equivalent
to the DBI action. A set of equations for general value of $p$ has been
written down, and it has been solved for $p$=2,3 and 4. This program
started out with the hope that there would be simplifications as compared
to the \DBI\ actions, but so far rather the contrary has happened. A
positive attitude would perhaps be that the correct set of auxiliary
fields still has to be found. As a consequence, the work described in
the following section, where the super- and $\k$-symmetric D-brane
actions are formulated, is performed entirely with the use of
\DBI\ kinetic actions.

\section{ACTIONS FOR SUPERSYMMETRIC D-BRANES}

This section reviews the work of refs.~\cite{CvGNW,CvGNSW}.
In the first section of this contribution (eqs.~(1) and (2)), we
described the action for a purely bosonic D-brane, containing
couplings to the massless bosonic fields of type \II\ string theories.
The most efficient way of writing down the full supersymmetric 
action is to replace
all background fields by the corresponding superfields, and view
the the bosonic D-brane world-volume as propagating in a superspace,
\ie\ the pull-backs are performed by $\*_iZ^M$, where $Z^M=(X^m,\t^\mu)$
are coordinates for the appropriate superspace. In such a formulation,
supersymmetry will be manifest. 
We will however perform a little manipulation on the expressions
of section 1. The reason for this is that the metric entering eq.~(1)
is the string metric, and not the Einstein metric. To this end we 
absorb a dilaton factor in the metric, and write the action as
\be
\vbox to34pt{
\hbox{$I=-\int d^{p+1}\x e^{-{p-3\/4}\phi}\sqrt{-\det(g_{ij}
	+e^{-{1\/2}\phi}\F_{ij})}$\hfill}
\vfill
\hbox{$\phantom{I=}+\int e^\F C\punkt$\hfill}
}
\ee
The relevant superspaces are those used in the superspace formulation
of the massless sectors of type \II A and \II B superstrings, \ie\
of type \II A and \II B supergravity. The type \II A superspace contains
two ten-dimensional Majorana-Weyl spinors of opposite chirality, while
in type \II B the chiralities are equal. In either case, we count to
a number of 32 real fermionic coordinates, which from a world-volume
point of view are propagating fields. This number is cut to half by the
fields equations. If we compare to the bosonic sector, containing
the 9-$p$ transverse components of the embedding and the $p$-1 of the
U(1) potential, there is an obvious mismatch of a factor two. The D-brane
world-volume field theory can not have a matching number of bosons and
fermions unless there is a mechanism that reduces the number of fermions
to eight. This is the r\^ole of $\k$-symmetry. It is the fermionic
gauge symmetry that gauges away half a spinor. The necessity of $\k$-symmetry
can also be understood from the observation that D-branes are BPS-saturated
states, breaking half the supersymmetry. A BPS-saturated multiplet
is a ``multiplet'' (the same size as a massless one), and $\k$-symmetry
leaves the correct number of fermions to generate such a multiplet.
The main issue in the formulation of supersymmetric D-branes is to prove
that the action (19) really possesses such a fermionic gauge symmetry.
Here, this construction will only be sketched, and the reader is referred
to refs.~\cite{CvGNW,CvGNSW} for the full details.

In order to make clear the details of the $\k$-symmetry, it is instructive
to begin with the flat superspace case. There, one has coordinates
$(X^a,\t^\a)$ (these letters are used throughout for inertial frame 
indices, and collectively denoted by $A$), 
and while a rigid supersymmetry transformation on the 
coordinates is
\be
\vbox to27pt{
\hbox{$\d_\e\t^{\a}=\e^{\a}\komma\phantom{\half}$\hfill}
\vfill
\hbox{$\d_\e X^a=-i(\bar\t\g^a\e)\komma\phantom{\half}$\hfill}
}
\ee
a $\k$-transformation takes the form
\be
\vbox to27pt{
\hbox{$\d_\e\t^{\a}=\k^{\a}\komma\phantom{\half}$\hfill}
\vfill
\hbox{$\d_\e X^a=i(\bar\t\g^a\k)\punkt\phantom{\half}$\hfill}
}
\ee
The only formal difference is the sign in the transformation of the
bosonic coordinate (and the fact that $\k$ is a local parameter).
Thus, while the generator of supersymmetry is the ordinary supersymmetry
generator, a generator of $\k$-symmetry is a covariant derivative,
and due to the fact that $\k$ is projected down to half a spinor,
only half a covariant derivative. The use of chiral superfields in
four dimensions is a way to eliminate this redundancy.

When we move to arbitrary curved superspace, the picture is equally
simple. In terms of the coordinates, the transformations read
\be
\d_\k Z^M=\k^\a {E_\a}^M\equiv\k^M\punkt
\ee
It is easily verified that this expression reduces to the ones above
for the choice of flat vielbein
\be
\vbox to27pt{
\hbox{${E_m}^a={\d_m}^a\komma
\phantom{{E_{\mu}}^a=i{\d_{\mu}}^{\a}(\g^\a)_{\a\b}\t^\nu{\d_\nu}^\b}
	\hskip-1.3cm{E_m}^{\a}=0\komma$\hfill}
\vfill
\hbox{${E_{\mu}}^a=i{\d_{\mu}}^{\a}(\g^a)_{\a\b}\t^\nu{\d_\nu}^\b\komma 
	\phantom{{E_m}^a={\d_m}^a}\hskip-1.3cm
	{E_{\mu}}^{\a}={\d_{\mu}}^{\a}\punkt$\hfill}
}
\ee
When investigating the transformation properties of the action (19), we
need the induced transformations of the background fields, and also
the transformation rule for the U(1) potential.
It follows straightforwardly from the coordinate transformations that the
induced transformations of any background field	$\Omega$ is
$\d_\k\Omega=\d_\k Z^M\*_M\Omega=\k^M\*_M\Omega$. When dealing with
pullbacks to the world-volume, we also have to take into account the
transformation of the the pullback vielbeins ${E_i}^M=\*_i Z^M$, and
obtain
\be
\vbox to27pt{
\hbox{$\d_\k g_{ij}=2{E_{(i}}^a{E_{j)}}^B \k^{\a}{T_{\a Ba}}
	\komma\phantom{\int}$\hfill}
\vfill
\hbox{$\d_\k C=\L_\k C=i_\k dC+di_\k C\komma\phantom{\int}$\hfill}
}
\ee
the latter equation valid for the pullback of any form ($T$ is the torsion
tensor).
The transformation of the U(1) potential is of course not a priori
given, it must be constructed so that the action is invariant.
It turns out that the correct transformation is
\be
\d_\k A=i_\k B\komma
\ee
which means that $\F=F-B$ transforms as
\be
\d_\k\F=i_\k dB=i_\k H\punkt
\ee
This also implies that the transformation of the WZ part of the action is
(modulo boundary terms, which will not be considered here)
\be
\d_\k I_\wz=\int e^\F i_\k R\komma
\ee
where the RR curvature cochain is 
\be
R=e^Bd(e^{-B}C)\punkt
\ee

In order to write down the action (19) more explicitly, we need to examine
which background fields enter this expression. These are the bosonic
fields of type \II\ supergravity, \ie\ the metric $g_{mn}$, the dilaton
$\phi$ and the antisymmetric tensor field $B_{mn}$ from the NS-NS sector
of the corresponding superstring theory, together with the antisymmetric
tensors of the RR sector. In a superspace formulation, these fields are
all superfields. They are not independent, but subject to constraints
that fix some of the components and relate other to each other, in order
to bring the total number of fields down to the physical content mentioned.
Here, we will display only part of these constraints, namely those
needed for proving $\k$-symmetry of the action. From the transformations
of the background fields above, it is clear that only components with 
at least one spinor index are affected. This allows us to consider only
fields of dimension 0 or $1/2$.

Before giving the constraints, it is necessary to digress on the 
relevant superspaces. In type \II A, the fermions are a pair of Majorana--Weyl
spinors of opposite chirality. They can be put together into a Majorana
spinor, which is identical to a Majorana spinor in eleven dimensions.
Besides the gamma matrices, there is an invariant matrix $\gel$ which
squares to $\one$ and anticommutes with the gamma matrices.
In type \II B, the two Majorana--Weyl spinors have the same chirality,
and can be considered as a (complex) Weyl spinor. However, this is not
convenient for our purposes. The division of the fields into the NS-NS
and RR sectors, which in this context corresponds to couplings in
the kinetic and WZ terms, respectively, roughly means splitting complex
fields in real and imaginary components. We therefore stay with a real
formulation, and in addition to the gamma matrices there are three invariant
real 2$\times$2-matrices $I$, $J$ and $K$ spanning the Lie algebra of
SL(2). Of these, $J$ and $K$ are symmetric and square to $\one$ 
and $I$ antisymmetric and squares to $-\one$. We also use
$IJ=K$ and all three matrices mutually anticommuting.

Our constraints are
\be
\vbox to390pt{
\vfill
\hbox{$\phantom{\hbox{\II A:\,\,\,\,}}{T_{\a\b}}^{c}
	=2i\g_{\a\b}^{c}\komma \qquad {T_{a\b}}^{c}=0\komma$\hfill}
\vfill
\hbox{$\hbox{\II A:\,\,\,\,}{T_{\a\b}}^\g=\genfrac{3}{2}{\d_{(\a}}^\g\LL_{\b)}
	+2{(\gel )_{(\a}}^\g(\gel \LL)_{\b)}$\hfill}
\vfill
\hbox{$\qquad\quad\,\,\,-\frac{2}{(\g_a)}_{\a\b}(\g^a\LL)^\g$\hfill}
\vfill
\hbox{$\qquad\quad\,\,\,+(\g_a\gel )_{\a\b}(\g^a\gel \LL)^\g$\hfill}
\vfill
\hbox{$\qquad\quad\,\,\, +\frac{4}{(\g_{ab})_{(\a}}^\g(\g^{ab}\LL)_{\b)}
	\komma$\hfill}
\vfill
\hbox{$\hbox{\II B:\,\,\,\,}{T_{\a\b}}^\c=-{(J)_{(\a}}^\c(J\LL)_{\b)}$\hfill}
\vfill
\hbox{$\qquad\quad\,\,\, +{(K)_{(\a}}^\c(K\LL)_{\b)}$\hfill}
\vfill
\hbox{$\qquad\quad\,\,\, +\frac{2}(\g_aJ)_{\a\b}(\g^aJ\LL)^\c\komma$\hfill}
\vfill
\hbox{$\qquad\quad\,\,\,-\frac{2}(\g_aK)_{\a\b}(\g^aK\LL)^\c\komma$\hfill}
\hbox{\hfill}
\hbox{$\hbox{\phantom{\II A:\,\,\,\,}} H_{\a\b\c}=0\komma$\hfill}
\vfill
\hbox{$\hbox{\II A:\,\,\,\,}H_{a\b\c}=-2i e^{\tfrac{2}\p}
	(\gel \g_{a})_{\b\c}\komma$\hfill}
\vfill
\hbox{$\hbox{\phantom{\II A:\,\,\,\,}}H_{ab\c}
	=e^{\tfrac{2}\p}(\g_{ab} \gel \LL)_{\c}\komma$\hfill}
\vfill
\hbox{$\hbox{\II B:\,\,\,\,}H_{a\b\c}=-2i e^{\tfrac{2}\p}
	( K\g_{a})_{\b\c}\komma$\hfill}
\vfill
\hbox{$\hbox{\phantom{\II A:\,\,\,\,}}H_{ab\c}
	=e^{\tfrac{2}\p}(\g_{ab}  K\LL)_{\c}\komma$\hfill}
\hbox{\hfill}
\hbox{$\hbox{\phantom{\II A:\,\,\,\,}} R_{(n)\a\b\c A_{1}...A_{n-3}}=0
	\komma$\hfill}
\vfill
\hbox{$\hbox{\II A:\,\,\,\,}R_{(n)a_{1}...a_{n-2}\a\b}$\hfill}
\vfill
\hbox{$\qquad\quad\,\,\,=2i\,e^{{n-5\/4}\p}
	(\g_{a_{1}...a_{n-2}}(\gel )^{{n\/2}})_{\a\b}\komma$\hfill}
\vfill
\hbox{$\hbox{\phantom{\II A:\,\,\,\,}}R_{(n)a_{1}...a_{n-1}\a}$\hfill}
\vfill
\hbox{$\qquad\quad\,\,\,=-\genfrac{{\xit n}-5}{2}e^{{n-5\/4}\p}
	(\g_{a_{1}...a_{n-1}}(-\gel )^{{n\/2}}\LL)_{\a}\komma$\hfill}
\vfill
\hbox{$\hbox{\II B:\,\,\,\,}R_{(n)a_{1}...a_{n-2}\a\b}$\hfill}
\vfill
\hbox{$\qquad\quad\,\,\,=2i\,e^{{n-5\/4}\p}
	(\g_{a_{1}...a_{n-2}}K^{{{n-1}\/2}}I)_{\a\b}\komma$\hfill}
\vfill
\hbox{$\hbox{\phantom{\II A:\,\,\,\,}}R_{(n)a_{1}...a_{n-1}\a}$\hfill}
\vfill
\hbox{$\qquad\quad\,\,\,=-\genfrac{{\xit n}-5}{2}e^{{n-5\/4}\p}
	(\g_{a_{1}...a_{n-1}}K^{{n-1\/2}}I\LL)_{\a}\punkt$\hfill}
\hbox{\hfill}
\hbox{$\hbox{\phantom{\II A:\,\,\,\,}}\LL_{\a}=\frac{2}\*_{\a}\phi\punkt$\hfill}
}
\ee

Constraints are formulated in terms of gauge-invariant quantities.
When a constraint is imposed on a set of curvatures, one has to check
that the Bianchi identities are still satisfied. This involves a number
of Fierz identities valid in the different superspaces, and these 
nor the Bianchi identities themselves will
be restated here. This is done in full in refs.~\cite{CvGNW,CvGNSW}.
For the original formulations of the type \II\ supergravities we refer
to refs.~\cite{HoweWest,Oerter}, where the superspace conventions
however are quite different from ours, that are streamlined for D-brane
calculations.

The constraints are to a certain extent a matter of convention.
One such convention is that the dimension-1/2 component ${T_{\a a}}^b$ 
of the torsion tensor is set to zero. The other dimension-1/2 component,
${T_{\a\b}}^\g$ can not consistently be set to zero, it is needed for a
full treatment of curved superspace. Note also that $\LL_\a$, the
superfield that contains the physical fermion as its leading component,
does not vanish in a general background, since it contains physical
fields in its higher components.
It should also be noted that the constraints put the supergravity 
theories on-shell.

The task is now to demonstrate that the action (19) is invariant under
$\k$-symmetry. This proof contains essentially two steps. The first one
consists of finding a projection operator that can be used to
project the transformation parameter to half its original number of
components, and the second one of performing the actual transformations
and to confirm that the contributions from the \DBI\ and \WZ\ terms cancel.

The projection operator $\Pi$ must have rank 16, since the superspaces
have 32 fermionic directions. We write it as $\Pi_\pm=\half(\one\pm\G)$,
where $\G$ is a traceless matrix that squares to $\one$. It then follows from
the fact that the eigenvalues of $\G$ are $\pm1$ that $\Pi$ is a projection
operator, and from the tracelessness that the number of positive and
negative eigenvalues are equal, so that the rank is correct.
We will use the $+$ sign, but this is just a matter of convention.

Some guidance can be obtained from the known properties of type I branes,
where no antisymmetric world-volume tensor is present, and $\G$ takes
the simple form
\be
\G={(-1)^{{1\/4}(p(p+1)+2)}\,\e^{i_1\ldots i_{p+1}}\/(p+1)!\sqrt{-\det g}}
	\,\,\g_{i_1\ldots i_{p+1}}\punkt
\ee
Here, world-volume gamma matrices are of course pullbacks of the space-time
ones. They depend on the fermionic coordinates even in the flat case,
through the projection vielbeins ${E_i}^M$.
One must be careful with the ``sign factor'' in this expression;
for certain values of $p$ it will be imaginary, which is unacceptable
when the spinors are real. In the D-brane case, this will be compensated
by appropriate factors of $I$ (which squares to $-\one$) or $\gel$ (which
anticommutes with gamma matrices). We note that this must essentially
be the leading part of the D-brane $\G$'s, obtained when $\F$ 
and the dilaton are set to zero. We also note the ``form character'' of $\G$,
that makes it possible to rewrite it as (disregarding the sign factor)
\be
d^{p+1}\xi\,\,\G\sim-{1\/\L}\,\,\g\komma
\ee
with $\g$ being the natural gamma matrix ($p$+1)-form 
\be
\g={1\/(p+1)!}\,d\xi^{i_{p+1}}\wedge\ldots\wedge d\xi^{i_1}
	\,\g_{i_1\ldots i_{p+1}}\punkt
\ee  
The projection operator $\Pi$ acts exactly as a chirality projection operator.

We can now state the form of the matrix $\G$ for D-branes. It is
\be
\vbox to 27pt{
\hbox{$d^{p+1}\x\,\G$\hfill}
\vfill
\hbox{$=-{e^{\tfrac{4}(p-3)\p}\/\L_{\dbi}}
	\exp\left(e^{-\tfrac{2}\p}\F\right)\wedge X Y|_{\rm vol}\komma$\hfill}
}
\ee
with 
\be
X=\bigoplus_{n}\g_{(2n+q)}P^{n+q}
\ee
and 
\be
\vbox to24pt{
\vfill
\hbox{$\hbox{\II A:\quad} P=\gel\komma\qquad Y=\id\komma\!\qquad q=1
	\komma$\hfill}
\vfill
\hbox{$\hbox{\II B:\quad} P=K\komma\qquad\,\; Y=I\komma\qquad q=0
	\punkt$\hfill}
}
\ee
In this expression, the ($p$+1)-form in the right hand side should be
singled out. To actually show that $\G$ squares to $\one$ is quite 
complicated, and we refer to ref.~\cite{CvGNSW} for the proof. The
tracelessness is trivial. The picture maybe becomes more clear from
an example. For the D3-brane, $\G$ takes the form
\be
\vbox to35pt{
\hbox{$\G={\e^{ijkl}\/\sqrt{-\hbox{\xrm det}(g+\F)}}$\hfill}
\vfill
\hbox{$\phantom{\G}\times\,\left(\frac{24}\g_{ijkl}I
	+\frac4\F_{ij}\g_{kl}J+\frac{8}\F_{ij}\F_{kl}I\right)\komma$\hfill}
}
\ee
and a similar pattern is followed by all the $\G$'s.

It now only remains to demonstrate the invariance of the action (19).
This is straightforward but tedious work. It of course makes use of the fact
stated above that all variations may be expressed in terms of the
dimension-0 and 1/2 background field components, and uses the solutions
of the Bianchi identities already put forward. The conclusion is that
we have proven that eq.~(19) correctly describes the dynamics of the
supersymmetric D-branes, provided they propagate in a background that 
solves the equations of motion for the effective action of the massless
fields of the appropriate superstring theory.

\section{CONCLUSIONS}

In this contribution, we have mainly discussed two aspects of actions
for D-branes, the possibility of simplifying them with the help of
an auxiliary non-propagating world-volume metric, and the formulation
of supersymmetric and $\k$-symmetric D-brane actions.

The investigation of the first of these issues resulted in partly
negative results --- although we were able to determine the form of
these actions for lower values of $p$, we were not able to detect a
clear pattern of the solutions possible to generalize to all $p$.
We also detected increasing complication as $p$ was increased.
Still, there should a reservation for the possibility that some additional
auxiliary fields might simplify the situation. The status of the
problem is not conclusive.

One of the main motivations behind this first investigation was the hope
that supersymmetrization might be made easier in such a framework.
We were however able to solve that second problem without resorting
to any auxiliary fields. The presentation of that solution was the 
other main issue addressed here. We were able to write down the general
action for a supersymmetric D-brane and demonstrate the mechanism of
$\k$-symmetry in a general supergravity background. Consistency of
the D-brane action required the supergravity equations of motion to be
satisfied (actually, the actions for higher $p$ require inclusion of
both the ordinary supergravity fields and their duals, which is only
possible on-shell).

There are of course many questions to address in the future in connection
to the present work. Some of the most interesting are related to 
M-theory in eleven dimensions \cite{Witten,Schwarz,Sen,Sezgin,Banks}, 
which in a fundamental and 
surprising way seems to be related to D-branes \cite{Banks}.
We would like to apply the techniques to the eleven-dimensional
fivebrane, whose action is not even known in a bosonic truncation. 
Closely connected
is the problem of treating configurations of multiple coinciding D-branes
\cite{WittenX}, 
which seems to go outside the realm of a classical space-time, into
a non-commutative geometry. This provides a concrete realization of
what for a long time generally has been suspected to be necessary
if one hopes to get hold of what non-perturbative string theory really is.

\def\nl{\\ \hbox to0pt{\hfill}}

\end{document}